\documentclass[aps,prb,twocolumn,showpacs]{revtex4}
\usepackage{graphicx}

\renewcommand{\vec}{\mathbf}

\begin{document}

\title{Finite Temperature Dynamics of the Spin 1/2 Bond Alternating Heisenberg
  Antiferromagnetic Chain}

\author{H. J. Mikeska and C. Luckmann}
\address{Institut~f\"ur~Theoretische~Physik, Universit\"at~Hannover, 
30167~Hannover, Germany}

\date{\today}

\begin{abstract}
  We present results for the dynamic structure factor of the $S=1/2$ bond
  alternating Heisenberg chain over a large range of frequencies and
  temperatures. Data are obtained from a numerical evaluation of thermal
  averages based on the calculation of all eigenvalues and eigenfunctions for
  chains of up to 20 spins. Interpretation is guided by the exact temperature
  dependence in the noninteracting dimer limit which remains qualitatively
  valid up to an interdimer exchange $\lambda \approx 0.5$. The temperature
  induced central peak around zero frequency is clearly identified and aspects
  of the crossover to spin diffusion in its variation from low to high
  temperatures are discussed. The one-magnon peak acquires an asymmetric
  shape with increasing temperature. The two-magnon peak is dominated by the
  $S=1$ bound state which remains well defined up to temperatures of the order
  of $J$. The variation with temperature and wavevector of the integrated
  intensity for one-magnon and two-magnon scattering and of the central peak 
  are discussed.

\end{abstract}

\medskip


\maketitle

\section{Introduction}
\label{sec:intro}

Low-dimensional gapped quantum antiferromagnets have received much interest in
recent years both experimentally and theoretically. They serve as model
substances allowing to investigate in detail the effects of quantum
fluctuations and to test theoretical models. A particularly simple class of
materials in this context consists of an assembly of dimers (two strongly
coupled spins 1/2) which interact sufficiently weakly to guarantee that the
dimer gap does not close. These materials are characterized by a disordered
singlet ground state and a finite spin gap to triplet excited states.
Materials of this type occur in nature as quasi one-dimensional (1D)
(Cu(NO$_3$)$_2\cdot$2.5H$_2$O) \cite{Tennant03}, quasi two-dimensional
(BaCuSi$_2$O$_6$) \cite{Jaime05} and three-dimensional ($\rm
KCuCl_3$ and $\rm TlCuCl_3$) \cite{Tanaka98,Cavadini00} compounds.

Systems of this type are characterized by both quantum and thermal
fluctuations: Thermal fluctuations are controlled by the temperature $T$ and
the quantum fluctuations of interest in dimer systems (i.e. those modifying
the basic quantum mechanics of isolated dimers) are governed by the interdimer
coupling. So far, most investigations have concentrated on the quantum
aspects, excluding thermal fluctuations by working at zero (theoretically)
resp. low (experimentally) temperature. The theoretical approach is then
restricted to the low-lying eigenstates of the quantum hamiltonian,
which are accessible to an approximate analytical treatment or to numerical
calculations. However, at finite temperature of the order of or even large
compared to the magnetic exchange energy, all levels participate and more
elaborate approximations resp. numerical approaches are necessary.

Recently increasing interest has developped in the finite temperature behavior
of such systems and experimental results for Cu(NO$_3$)$_2\cdot$2.5H$_2$O
\cite{Tennant03,Notbohm06} and $\rm TlCuCl_3$ \cite{Ruegg05} demonstrate that
an understanding of the interplay of thermal and quantum fluctuations requires
additional work on the theoretical and numerical side. Our aim here is to
investigate for the system of weakly coupled dimers described above the
dynamical structure factor over a wide range of frequencies and temperatures.

To explore from numerical calculations the content of a given hamiltonian for
finite temperatures is more involved than for $T=0$ since the Lanczos approach
which allows to treat reasonably large systems (up to 36 spins $S=1/2$) does
not give more than a few low-lying energy levels. Instead, the direct approach
to dynamic properties at finite temperatures requires full exact
diagonalization (FED) to obtain energies and wave functions; this has been
done so far for the XXZ S=1/2 chain \cite{Fabricius97} and the frustrated
S=1/2 Heisenberg chain \cite{Fabricius98} up to 16 spins. If FED is restricted
to energy levels only, thermodynamic properties such as the specific heat,
susceptibility and structure factor $S(q)$ are obtained; this has been done
for a number of dimer type systems such as the dimerized and frustrated
$S=1/2$ chain \cite{Uhrig01} and the two-leg ladder \cite{Uhrig03a}.  These
numerical approaches to finite temperatures have been supplemented by
approximate analytical approaches such as including thermal occupation factors
in the mean field approach \cite{Troyer94,Ruegg05} and a strong coupling field
theoretic apporach \cite{Damle98}.

We describe the one dimensional (1D) $S=1/2$ chain with alternating isotropic
antiferromagnetic nearest-neigbour exchange (the bond alternating Heisenberg
chain, BAHC) as a system of $N/2$ unit cells with two spins each and two
exchange constants, the intradimer exchange $J$ and the interdimer exchange
$\lambda J$, using the following hamiltonian:
\begin{equation}
\label{eq:hambahc}
\mathcal H = J \ \sum_{n=1}^{N/2} 
             \left( {\mathbf S}_{n,1} \cdot {\mathbf S}_{n,2}
             + \lambda \ {\mathbf S}_{n,2} \cdot {\mathbf S}_{n+1,1} \right),
\end{equation}
We assume $J>0$ and apply periodic boundary conditions. For $\lambda=0$
the ground state of the system consists of singlets on the intradimer bonds
$(n,1) - (n,2)$.  These local singlets can be excited to triplets which
develop into a band of gapped excitations when switching on $\lambda$. At
higher energies multiparticle excitations dominate the spectrum. In the limit
$\lambda=1$ we arrive at the well known Heisenberg antiferromagnetic chain
(HAF) with pairs of $S=1/2$ spinons as lowest gapless excitations. Other
related models are described in Ref.~\onlinecite{Brehmer96}.

The dynamics of such a system is most appropriately discussed in terms of the
dynamic structure factor, i.e. the Fourier transform of the time dependent
spin correlation function which, apart from well known prefactors gives the
spectral weight measured as the magnetic inelastic neutron scattering (INS)
cross section \cite{Squ78}. The dynamic structure factor of the BAHC at low
temperatures ($T \ll J$) is characterized by a peak due to singlet-triplet
transitions as most prominent feature and continua of multiparticle
excitations at higher energies. From extended work on the $T=0$ quantum
mechanics of this system
(Ref.~\cite{Uhrig96,Barnes99,Trebst00,Hamer03,MMuller03} and references
therein) it is known that finite interdimer coupling $\lambda$ leads to
considerable modifications in energies and transition strengths for one and
two magnon processes as well as to the emergence of bound states below (and
above) the two magnon band. Here we present results from FED for up to
20 spins for the longitudinal dynamic structure factor (DSF) $S^{zz}(\vec{q},
\omega)$ for a wide range of temperatures, varying from $T \ll J$ to $T \gg
J$. Our discussion will concentrate on the variations of the one and two
magnon lineshapes with temperature (with particular emphasis on the bound
state) and on the temperature dependence of the contributions from the
temperature induced 'central peak' at $\omega \approx 0$ which is due to
intraband transitions. This peak is the signature of spin diffusion in the
classical limit; its temperature dependence has not been discussed so far in a
microscopic context, but it is clearly seen in recent INS experiments
\cite{Notbohm06}.

For $N$ spins localized on sites $\vec{x}_i$ on a 1D lattice the dynamical
structure factor is defined as
\begin{eqnarray} 
\label{eq:dynamicsf}
  S^{\alpha\beta}(\vec{q}, \omega) &=& \frac{1}{2 \pi N} \ \sum_{i,j} \
  \int_{-\infty}^{\infty} \!\!dt\ e^{i \, (\vec{q} \, (\vec{x}_i-\vec{x}_j)
       \ - \, \omega t)} \nonumber \\
  &&  \langle \ S^{\alpha}_i(t) \ S^{\beta}_j(t=0) \ \rangle
\end{eqnarray}
The superscripts $\alpha, \beta$ denote the spin components and the brackets
$\langle\cdots\rangle$ thermal expectation values (which for $T=0$ reduce to
groundstate expectation values $\langle 0 | \cdots |0\rangle$).  For the
isotropic hamiltonian (no magnetic field) of Eq.~(\ref{eq:hambahc}) the
dynamic structure factor is diagonal with three equal components. It is
therefore sufficient to restrict the discussion in the following to one
component, $S^{zz}(\vec{q}, \omega)$. In the presence of a magnetic field, a
case to be treated in subsequent work, we will have to distinguish between 
$S^{zz}(\vec q, \omega )$ and $S^{+-}(\vec q, \omega )$.

The remainder of this article is organized as follows: In section
\ref{sec:limits} we review and discuss simple limiting situations and our
numerical approach, in section \ref{sec:results} we present results for the
BAHC with $\lambda$ up to 0.3 and temperatures up to $4 \, J$ and in the final
section \ref{sec:conclusion} we discuss the range of validity of our approach
and give our conclusions.

\section{Limiting cases and numerical approach}
\label{sec:limits}

The BAHC has lattice sites $\vec{x}_i = n \, \vec{b} + p \, \vec{d}/2$, $n = 1
...  N/2, \, p = \pm 1$. Here $\vec{d}$ is the intradimer distance, the
direction of $\vec{d}$ in general differs from the chain direction given by
$\vec{b}$.  At finite temperatures the general expression for
$S^{zz}(\vec{q},\!\omega)$ is written as

\begin{eqnarray} 
&S^{zz}(\vec{q},\omega)& = \sum_i \frac{e^{-\beta E_i}}{Z} \cdot 
                                                 \nonumber \\[2mm] 
   && \sum_k \ \vert \langle \psi_k \vert \ S^z_{\vec{q}} \ 
     \vert \psi_i \rangle \vert^2 \ \delta(\omega - (\omega_k - \omega_i))
\end{eqnarray} 
where $Z$ is the partition function and we have introduced the Fourier 
transformed spin operators
\begin{equation} 
S^z_{\vec{q}} = \frac{1}{\sqrt{N}} \sum_{n=0}^{N/2 -1} \sum_{p=\pm 1}
                e^{i \vec{q} \ (n \vec{b} + p \vec{d}/2)} \ S^z_{n,p}.
\end{equation} 

We begin by reviewing results for the two limiting cases $T=0$ (pure quantum
effects) and $\lambda =0$ (pure thermal fluctuations). The BAHC at $T=0$ with
its singlet-triplet gap is one of the simplest systems to study strong quantum
fluctuations in numerical and analytical approaches. This has been done by
many different methods, among them the random phase approximation
\cite{Uhrig96} and series expansions in the coupling strength $\lambda$
\cite{Barnes99,Trebst00,Hamer03,MMuller03}. The application of these methods
has concentrated on two types of low-lying excitations, i.e.~eigenstates of
the hamiltonian, above the singlet ground state: (i) The basic quantum
excitation is the first excited triplet, i.e.~the propagating one magnon
excitation (one excited dimer) (ii) two magnon excitations which form a
continuum with the possibility of bound states. These states are decorated by
admixtures of multimagnon excitations which become more and more important
when the interdimer coupling $\lambda$ increases. These methods have
established the existence of bound states of two excited dimers below the two
magnon band, two singlets and two triplets \cite{Trebst00}, for wave vectors
close to $\pi$ and energies and transition matrix elements for the one and
two magnon bands have been obtained up to $13^{th}$ order \cite{Hamer03} from
series expansions. We remark that the analogous calculation including 3D
interactions has only been done up to $3^{rd}$ order \cite{MMuller03}. These
results are in excellent agreement with experiments at low temperatures.

In the limit of vanishing coupling, $\lambda=0$, on the other hand, the BAHC
reduces to an assembly of noninteracting dimers and its exact dynamical
structure factor can be easily calculated for all temperatures. In spite of
the simplicity of the calculation the result is instructive as a guide to the
general case. The full result for $S^{zz}(\vec q, \! \omega)$ in this limit is
\begin{samepage}
\begin{eqnarray}
\label{eq:dynamicsfzero}
&&S^{zz}_{\lambda=0}(\vec q, \omega) = 
        \frac{1}{4} \frac{1}{1 + 3 e^{-\beta J}}
\left\{ 2 (1 + \cos \vec q \cdot \vec d)  
    e^{- \beta J} \,\delta(\omega) \right. \nonumber \\[2mm]
 &&  + \left. (1 - \cos \vec q \cdot \vec d) 
        \left( \delta (\omega -J) 
  + e^{- \beta J} \delta(\omega + J) \right) \right\} 
\end{eqnarray}
\end{samepage}
This expression is valid for any number of spins $N$. The wavevector $\vec q$
is in chain direction and $q$ takes only the $N/2$ discrete values $q = (2\pi
m)/(N b/2), \ m=1, 2, \dots N/2$ of the lattice of dimers (lattice constant
$\vec b$).  Eq.~(\ref{eq:dynamicsfzero}) shows the two types of contributions
which survive the limit $\lambda \to 0$, i.e. which are present in the
noninteracting dimer limit: (i) one magnon excitations with energy $\omega =
\pm J$ (related in strength by the detailed balance factor ${\rm e}^{- \beta
  J}$); (ii) a contribution at $\omega =0$ which results from transitions
(diagonal terms for $S^{zz}$) within the excited dimer triplet. It therefore
carries a factor ${\rm e}^{- \beta J}$ and will develop into a 'central peak'
for $\lambda \ne 0$. Since all transition processes in this limit are
localized, the only nontrivial wavevector dependence results from form factors
related to the internal dimer structure: the intradimer distance $\vec d$
determines the 'dimer structure factor' $(1 - \cos \vec q \cdot \vec d)$ and
the corresponding factor $(1 + \cos \vec q \cdot \vec d)$ for the central
contribution.

The result of Eq.~(\ref{eq:dynamicsfzero}) is independent of $N$ since for
independent dimers, $\lambda = 0$, the correlation length vanishes and the
exact result is obtained already for $N=1$. Increasing $N$ only changes the
number of wave vectors $\vec q$ and their positions. This guarantees the
extensivity of the sum rule $\sum_{\vec q}\int d\omega S^{zz}(\vec q,\omega) =
N/4$ \cite{footnote01}. The absolute intensity of the one 'magnon' peak at
e.g. $q = 2 \pi$ develops from 1/2 at $T=0$ to 1/8 at $T \to \infty$. We note
that the various contributions to this intensity from states with a different
number of excited dimers depend on $N$ as can be identified easily from
Eq.~(\ref{eq:dynamicsfzero}): The prefactor in Eq.~(\ref{eq:dynamicsfzero}),
\begin{equation}
\label{weightsfzero}
\frac{1}{1 + 3 \, {\rm e}^{-\beta J}} =
 \frac{(1 + 3 \, {\rm e}^{-\beta J})^{N/2 -1}}
                       {(1 + 3 \, {\rm e}^{-\beta J})^{N/2}}
\end{equation}
originates from the partition function in the denominator and the phase space
factor resulting from excited dimers which do not participate in the 
transition in the numerator.
The expansion of the numerator thus gives as a series the relative strength of
the various contributions with energy $J$ starting from a state with $M$
excited dimers: The contribution of states with $M$ excited dimers is
proportional to ${N/2 \choose M} (3 \, {\rm e}^{-\beta J})^{M}$. For $T \to
\infty$  this reduces to ${N/2 \choose M} \ 3^M$, the
degeneracy of states with $M$ excited dimers. From this follows that the
various contributions to the dynamic structure factor from states with $M$
excited dimers have a well defined maximum (for $N \gg 1$) at $M = M_0 = z N$
with $z = 3/({\rm e}^{\beta J} +3)$. This means $M_0 \approx 0.524 N$ for
$T=J$ and $M_0 = 0.75 N$ for $T \to \infty$.
 
It should be noted that even for $T \to \infty$ the result of Eq.
(\ref{eq:dynamicsfzero}) is a quantum mechanical result reflecting the
discrete energy spectrum of a $S=1/2$ dimer. It does not agree with
the classical limit which is defined by the limit $\hbar \to 0$,
i.e.~for spins which are classical vectors or alternatively in the
limit $S \to \infty$. From the latter approach follows that the
frequency regime $0 < \omega < 2 J S$ fills continuously with
excitation strength in the classical limit.  The result of
Eq.~(\ref{eq:dynamicsfzero}) can clearly also be generalized to
include an external magnetic field, a case which will be treated in a
forthcoming paper.

\begin{figure}
\centering
\raisebox{0mm}{\includegraphics[height=75mm]{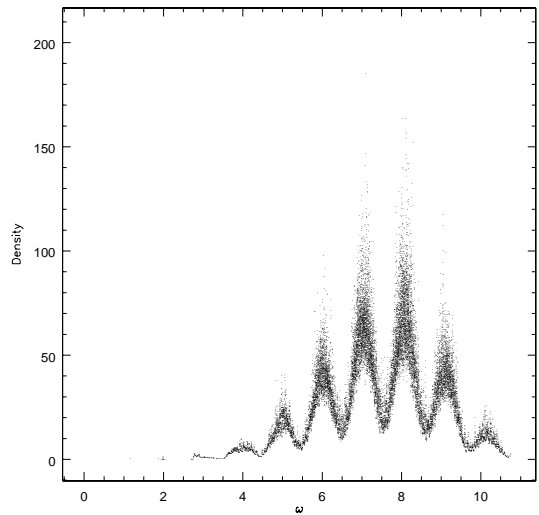}}
\caption{Density of states for $N=20, \ \lambda = 0.3, \ q = \pi$. Each
  point in the figure represents one multiplet. 
}
\label{fig:density} 
\end{figure}

Allowing for finite coupling $\lambda$ several features appear: (i) the
contributions at $\omega=0$ and $\omega = \pm J$ acquire a finite width owing
to the dispersion of the excitations, (ii) additional contributions at $\omega
\approx \pm p J$ appear in increasing order in $\lambda$, (iii) the intradimer
form factors get modified \cite{MMuller03}. For small coupling $\lambda \ll 1$,
two essential characteristics of the noninteracting limit survive
qualitatively as shown for $N=20, \lambda = 0.3$ in Fig.~\ref{fig:density}:
The correlation length remains small and the density of states continues to
show a well defined sequence of peaks with maxima at $E \approx p J$ up to the
upper bound of the spectrum and with widths of the order of $\lambda$
independent of energy. Thus the largest contributions to the transitions
$\omega \approx J \ (\omega \approx 2J)$, which start at energies $E \approx
M_0 \ J$ will be included for $N \ge 16 \ (N \ge 20)$ even for $T \to \infty$.
We conclude that these system sizes will be sufficient to obtain reasonable
results for the BAHC structure factor at moderate $\lambda$ and all
temperatures in its most interesting frequency regimes.

For our numerical calculations we have used the Householder algorithm to
obtain all eigenvalues and eigenvectors of the hamiltonian in the
subspace of constant $S^z_{\rm tot}$ and wavevector. Calculations were
performed on workstations with Pentium IV processors at the ITP,
Hannover University and on the JUMP supercomputer at NIC
J\"ulich. Calculations for $N=16$ are comfortably to perform (matrix
dimension is about 2000, required memory about 100MB). A few
calculations have been performed for $N=20$ for one value of $\lambda$
and the wavevector and frequency regime where the bound triplet two magnon
state is found; here memory (a few GB) and required CPU time are at
the limit what can be reasonably managed at present.

\section{Results and discussion}
\label{sec:results}

\begin{figure}
\centering
\raisebox{0mm}{\includegraphics[width=75mm]{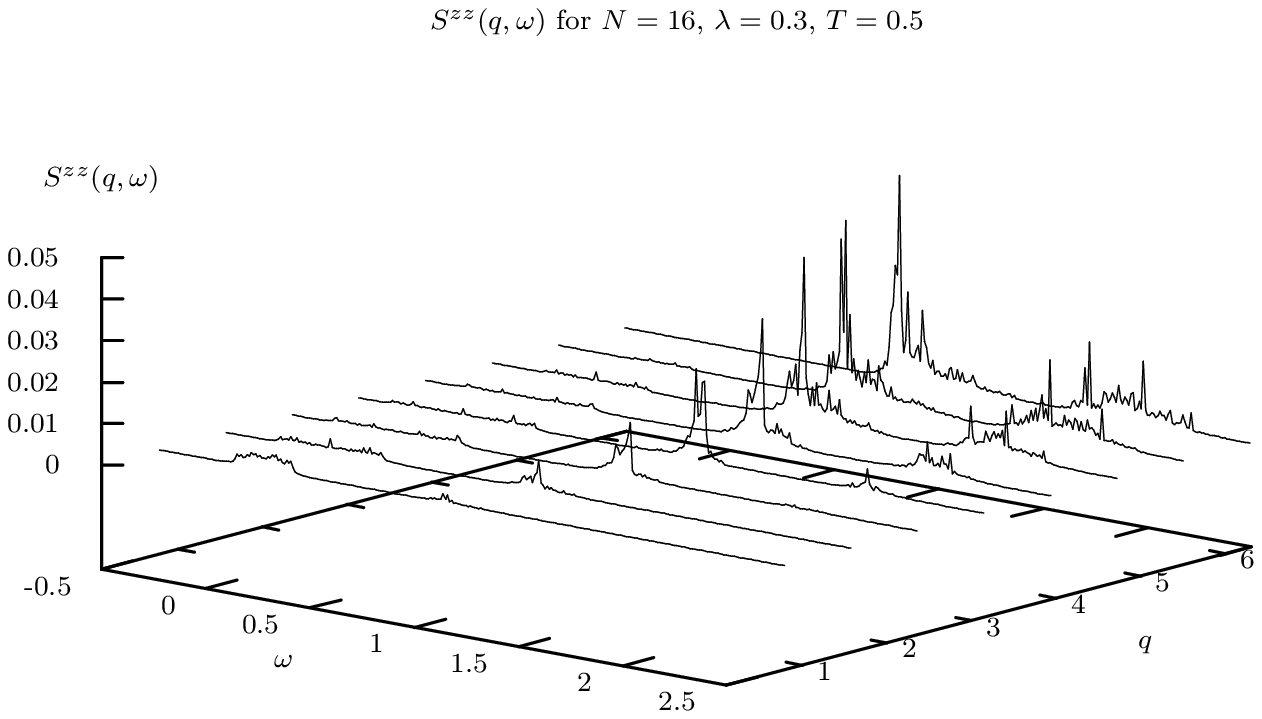}}\\[10mm]
\raisebox{0mm}{\includegraphics[width=75mm]{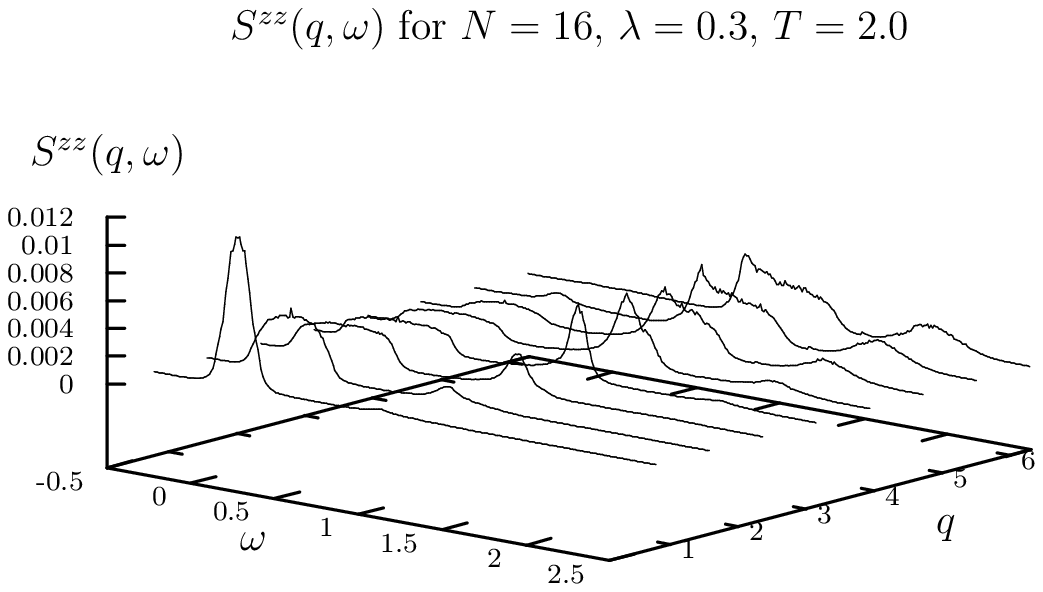}}
\caption{Overall picture of the dynamic structure factor $S^{zz}(q, \omega)$ 
for $N=16, \lambda = 0.3$, (a) $T=0.5$, (b) $T=2.0$. The intensity of
the two magnon peaks is enlarged by the factor 20.} 
\label{fig:dsfoverall} 
\end{figure}

\begin{figure}
\centering
\raisebox{0mm}{\includegraphics[width=85mm]{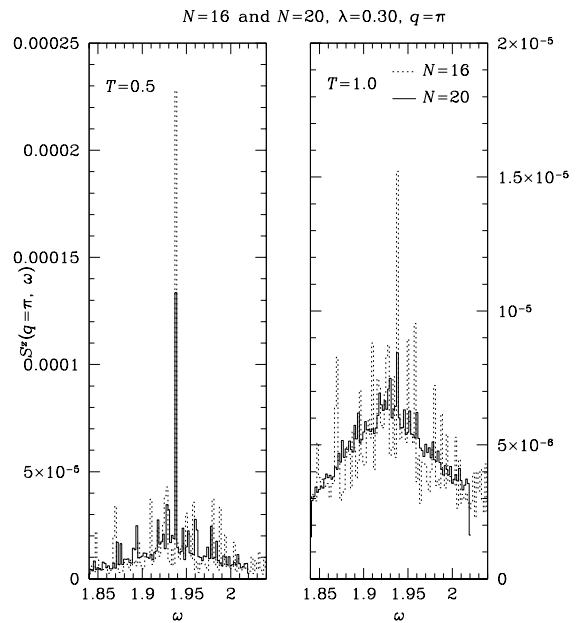}}
\caption{(Color online) Two magnon peak for $q=\pi, \lambda = 0.3$ and $N=16$ 
and $N=20$ (for the frequency interval 
$1.84 <\omega < 2.02$) in comparison: (a) $T=0.5$ and (b) $T=1$.
Here, narrow frequency bins, $\Delta \omega = 0.002$, are used.}
\label{fig:twomagnons20}
\end{figure}

In this section we present our results for $S^{zz}(\vec q, \omega)$. We cover
the full range of temperatures (in units of $J$), $T = 0.1 .... T = 4$ (the
latter being representative for $T \to \infty$) and wave vectors $q$ (in units
of $b^{-1}$) from 0 to $2\pi$. As standard values we use $N=16$ and $\lambda =
0.3$. We find that the variation with $\lambda$ for smaller values can be
safely found from interpolating the results presented here. The number of
spins strongly influences the level density and the discreteness of reciprocal
space; however, the positions of levels which occur at identical wavevectors
for different $N$ (e.g.~the one magnon state and the $S=1$ bound state below
the two magnon band at wavevector $\pi$) do not change with $N$ going from
$N=12$ to $N=24$ (values for $N=24$ were obtained from Lanczos
diagonalization). In this work we consider the simplest case $\vec d = \vec
b/2$. Then $d$ is the distance between all spins and it is sufficient to
specify $q = \vert \vec q \vert$ in the following. $d = b/2$ implies that the
effects from the dimer form factors are particularly simple: in lowest orders
the central contribution vanishes for $q=2\pi$ and the one magnon contribution
for $q=0$.  Temperatures and frequencies are given in units of $J$ throughout
this section.  The complete variation of $S^{zz}(q, \omega)$ with wavevector
and frequency for $N=16$ and $\lambda = 0.3 $ is shown in
Fig.~\ref{fig:dsfoverall} for two temperatures, $T=0.5$ and $T=2$. The small
contribution of the two-magnon excitations has been enlarged by a factor 20.
The remaining figures display the details of the influence of temperature on
the various interesting aspects of the spectra. The scale of $S^{zz}(q,
\omega)$ is set by the value for the one-magnon peak for noninteracting dimers
at $T=0$ and $q = 2\pi$, i.e.  1/2 in the units of Eq.~(\ref{eq:dynamicsf}).
For all presentations the transitions strengths have been added up within
frequency bins of mostly $\Delta \omega = 0.01$ ($0.002$ in
Fig.~\ref{fig:twomagnons20}).

\begin{figure}
\centering
\raisebox{0mm}{\includegraphics[width=85mm]{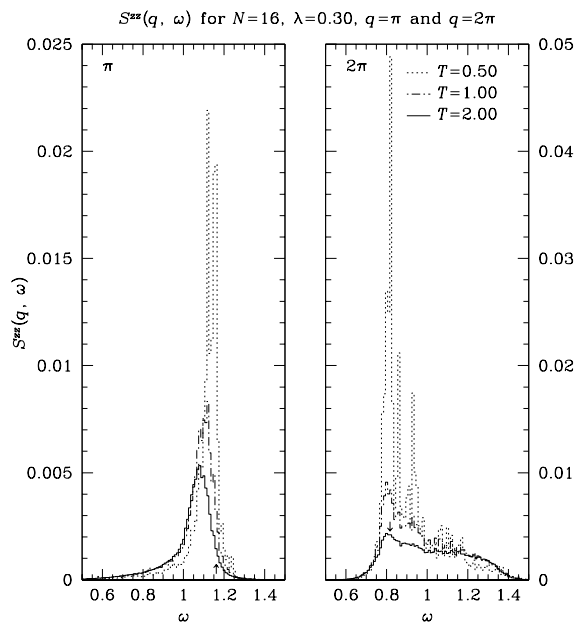}}
\caption{(Color online) One-magnon peak for $N=16, \ \lambda = 0.3$: 
(a) $q=\pi$ and (b) $q = 2\pi$. Temperatures are $T= 0.5$ (dotted lines), 
$T= 1.0$ (dash-dotted lines), $T=2.0$ (full lines). The arrows indicate the 
frequencies of the one magnon transition at $T=0$.}
\label{fig:onemagnon}
\end{figure}

Fig.~\ref{fig:onemagnon} illustrates the evolution with temperature of the
one-magnon peak for $N=16$ and $\lambda = 0.3$. With increase in temperature,
the peaks extend over a nearly constant range in frequency, although the
decrease of the maximum of intensity formally implies an increase in the width
at half maximum. For both $q=\pi$ and $2 \pi$ the position of the maximum
shifts to lower frequencies, but the peak develops a marked asymmetry with
more intensity on the high frequency side. This effect is most pronounced for
$q=2\pi$, the wave vector of the energy gap (and may not be the case at all
for small values of $q$, see Fig.~\ref{fig:dsfoverall}). Thus the detailed
description of the lineshape provides an understanding for the puzzling
observation that the gap energy seems to increases with temperature
\cite{Ruegg05} (comparable to an analogous observation in the Haldane chain
\cite{Jolicoeur94}). It would be interesting to compare the lineshape to the
result of the theoretical approach of Ref.~\onlinecite{Damle98}. However, with
$N=16$, a continuous lineshape results only for temperatures above the energy
gap and is thus in the nonuniversal approach of the approach of
Ref.~\onlinecite{Damle98}. When we look at the microscopic origin of the peak
broadening, we find that e.g.~at $T=1.0$ the basic transition from the ground
state to the one magnon excitation at $q=\pi$ is responsible for 30\% of the
weight at that frequency and that for the neighboring frequencies transitions
starting from the one magnon band contribute about 10\% of the total weight,
whereas the by far largest part of the intensity originates from transitions
starting at states with two or more excited dimers.

\begin{figure}
\centering
\raisebox{0mm}{\includegraphics[width=85mm]{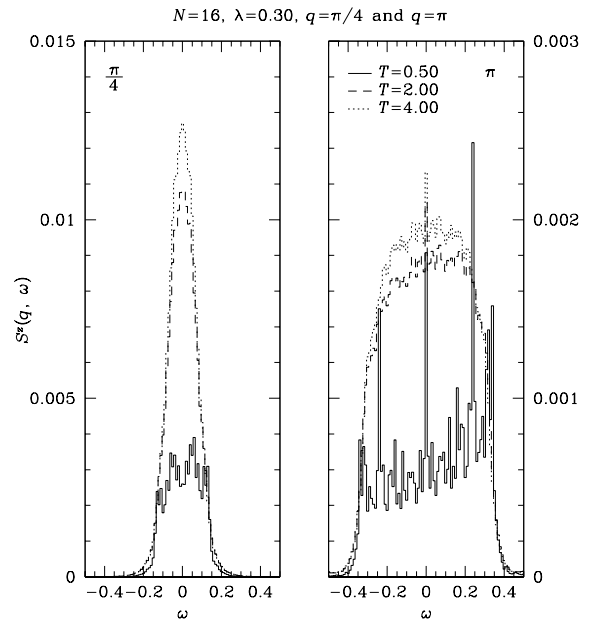}}
\caption{(Color online) Central peak for $N=16, \ \lambda = 0.3$: 
variation with temperature up to $T=4$ for (a) $q=\pi/4$ and (b) $q=\pi$}
\label{fig:centralT}
\end{figure}

\begin{figure}
\centering
\raisebox{0mm}{\includegraphics[width=75mm]{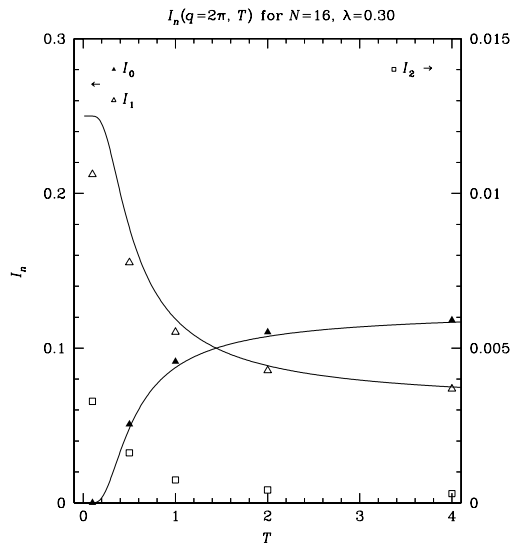}}
\caption{Temperature dependence of the 'exclusive structure factors' 
$I_n(q, T)$ for 
$N=16, \lambda = 0.3$ and $q=\pi$. $n = 0, 1$ (left scale, lines
give the result for $\lambda = 0$ from Eq.~(\ref{eq:dynamicsfzero}))
and $n =2$ (right scale)}
\label{fig:exclusivedsfT}
\end{figure}

Fig.~\ref{fig:centralT} illustrates 
for $N=16, \ q = \pi/4$ and $\pi$ the evolution of the central peak with
temperature 
(for $\lambda = 0.3$). At low $T$ and $\lambda \ll 1$ the shape 
of the central peak is dominated
by transitions inside the weakly interacting one magnon band with
$q-$independent matrix elements. The physical process may be thought of as
an external probe accelerating a thermally populated excitation and is thus
similar to well known processes in soliton bearing 1D magnets \cite{Villain75}.
As in these models, the limiting form of the structure factor is
\begin{equation}
\label{eq:shape}
S^{zz}_{(0)}(q,\omega) \propto e^{-\beta J} \ 
               \frac{1}{\sqrt{1 - \omega^2/\omega_m(q)^2}} \ 
               \Theta(\omega_m^2(q) - \omega^2)    
\end{equation}
with $\omega_m(q) = \lambda \sin (q/2)$ just from phase space effects (a small
variation with temperature resulting from the dispersion has been neglected).
Numerically we find that the cutoff of the central peak remains localized at
$\omega \approx \omega_m(q)$ for all wavevectors and temperatures (apart from
corrections of ${\cal O}(\lambda^2))$. The lineshape, however, cannot be
expected to be reproduced since only a few discrete transitions in the one
magnon band are available for $N=16$ and dominate the spectrum at the lowest
temperatures.  Nevertheless, for $q = \pi, T=0.5$ the inverted lineshape of
Eq.~(\ref{eq:shape}) starts to become visible even with this restriction.
Between medium ($T=0.5$) and high ($T=4$) temperatures a crossover of the
lineshape from square like to Gaussian is observed for small wavevectors.  In
view of the experimental \cite{Takigawa96} and theoretical
\cite{Zotos04,Damle05} discussion of spin diffusion in gapped 1D magnets we
have also made a rough analysis of the linewidth variation with wavevector. We
find that the $T=4$ spectra fit surprisingly well to a Gaussian (but not to a
Lorentzian) and the linewidth increases by a factor of about 5 upon doubling
the wavevector from $q=\pi/4$ to $\pi/2$. This indicates that the crossover to
spin diffusion like behavior governs the behavior of the DSF in the regime
considered here. The decrease of the overall intensity with temperature due to
the thermal occupation factor will be discussed below, see
Fig.~\ref{fig:exclusivedsfT}.

\begin{figure}
\centering
\raisebox{0mm}{\includegraphics[width=85mm]{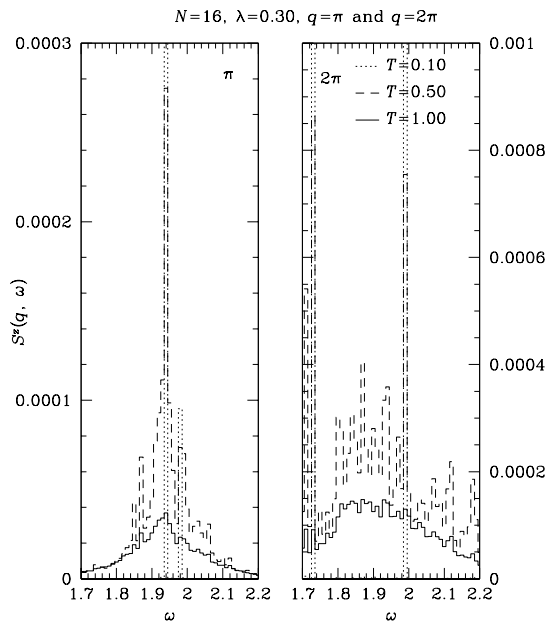}}
\caption{(Color online) Two magnon peak for $N=16, \lambda = 0.3$ and 
different 
temperatures (symbols as in Fig.~\ref{fig:onemagnon}): (a) $q=\pi$ (the $S=1, 
T=0.1$ bound state contribution is at 0.00318 outside the frame), (b) 
$q=2 \pi$ (the two largest $T=0.1$ contributions are outside the frame
at 0.0109 for $\omega 
\approx 1.73$ and at 0.00961 for $\omega \approx 1.99$)}
\label{fig:twomagnons16}
\end{figure}

Fig.~\ref{fig:twomagnons16} shows the two-magnon peak at $q = \pi$ (where the
bound states are most clearly visible) and $q = 2 \pi$ for $N=16, \ \lambda =
0.3$ \ in the temperature range $T=0.5 \dots 2$ (note the enhancement in
frequency and intensity scales). For $q=\pi$ the low temperature spectra are
entirely dominated by the $S=1$ bound state at $\omega \approx 1.938...$,
whereas the continuum (which at $T=0$ is smaller by a factor of $\lambda^2$ in
intensity \cite{Hamer03}) plays no significant role. The bound state remains
clearly visible up to $T=1$ and then disappears in parallel with rapid
decrease in the integrated intensity of the two-magnon peak for temperatures
above $T=1$ (see also Fig.~\ref{fig:exclusivedsfT} below). This reflects the
fact that dimers become independent of each other with increasing temperature
such that the correlations between spins in two different dimers required for
a finite two-magnon peak disappear. In Fig.~\ref{fig:twomagnons20} we present
a comparison between results for $N=16$ and $N=20$ (available only for $q=\pi$
and the limited frequency range $1.84 < \omega < 2.02$). In going from $N=16$
to $N=20$ some improvement is obtained, the continuum becomes smoother and the
bound state less dominant, but the main characteristics are unchanged. Thus
the increase in $N$ is not really crucial. The spectra invoke the impression
that at intermediate temperatures additional transitions (in particular the
$S=0$ state at $\omega \approx 1.868$ which could be reached from a thermally
excited $S=1$ state) become visible. This, however, is misleading as the
comparison between the data for $N=16$ and $N=20$ shows: A change in the
number of spins leads to a different set of allowed wave vectors and thus,
from the accompanying change in the initial and final energies, to trivially
different excitation frequencies (although the energies at e.g. $q=\pi$
are remarkably independent of $N$).  For $q=2\pi$ there is no bound state; the
two discrete transitions at $T \approx 0$ in Fig.~\ref{fig:twomagnons16}b are
the remainder of the continuum for the limited number of spins 16.

For a more global analysis we show in Fig.~\ref{fig:exclusivedsfT} the
intensities $I_n(q=\pi, T)$ obtained from integrating $S^{zz}(q=\pi, \omega)$
over frequency intervals $(n-\frac{1}{2}) < \omega < (n+\frac{1}{2})$. The
results for $q=\pi$ are representative also for the other $q-$values. These
'exclusive structure factors' show most clearly the relevance of quantum
fluctuations with increasing interdimer coupling $\lambda$ at low
temperatures. Quantum fluctuations are most important for the one-magnon peak
$I_1$ at low temperatures. $I_1(q=\pi,T)$ comes close to its classical limit
independent of $\lambda$ at $T \approx 2$, whereas the temperature dependence
of $I_0(q=\pi,T)$ follows closely the $\lambda = 0$ result of
Eq.~(\ref{eq:dynamicsfzero}) for all temperatures. The temperature variation
of $I_2(q=\pi,T)$ shows clearly that an observation of the two-magnon peak
requires low temperatures $T < 1$. Transitions corresponding to exciting more
than two dimers are extremely small in magnitude: A typical number is $I_3(q=
2\pi, T=0.5) \approx 2 \cdot 10^{-5}$.  From further calculations of
integrated intensities we conclude that the number of spins is not very
important for the overall features: $N=12$ gives nearly undistinguishable
results.

\begin{figure}
\centering
\raisebox{0mm}{\includegraphics[width=75mm]{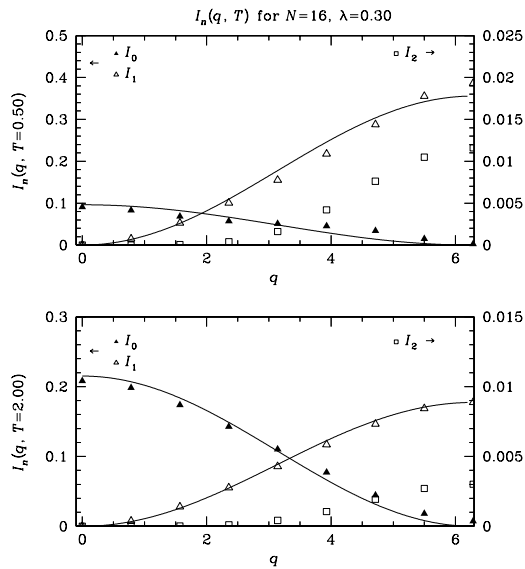}}
\caption{Wave vector dependence of the 'exclusive structure factors' 
$I_n(q, T)$ for (a) $T=0.5$ adn $T=2.0$. 
Lines for $n = 0, 1$
give the result for $\lambda = 0$ from Eq.~(\ref{eq:dynamicsfzero}).
Units for $I_2$ (right scale) are enhanced by a factor of 20.}
\label{fig:exclusivedsfq}
\end{figure}

Fig.~\ref{fig:exclusivedsfq} shows the influence of temperature and
interdimer interaction on the dimer structure factor $(1- \cos qd)$
for the one magnon intensity and the corresponding factor $(1 + \cos
qd)$ for the central peak. In the analysis of experiments this factor
is usually considered as a prefactor independent of $\lambda$ and
$\omega$, although it is clear already from the zero temperature
analysis \cite{MMuller03} that corrections are present in higher
orders in $\lambda$.  Fig.~\ref{fig:exclusivedsfq} shows an overall
agreement between the numerical results for $\lambda = 0.3$ and those
of the noninteracting limit although closer inspection reveals
significant differences: The ratio $I_1(q=2\pi, T=0.5)/I_1(q=\pi, T=0.5)$
e.g.~is about 2.5 for $\lambda = 0.3$ compared to 2 for $\lambda = 0$.
Fig.~\ref{fig:exclusivedsfq} also displays the wavevector dependence
of the two magnon peak which is absent in the noninteracting limit:
The intensity of this peak depends strongly on wavevector and wave
vectors $q > \pi$ are most favorable for an observation of these
processes.

\section{Conclusions}
\label{sec:conclusion}

We have calculated the dynamic structure factor of the bond
alternating Heisenberg chain based on full exact diagonalization for
chains with up to 20 spins 1/2. This allows us to obtain results at
finite temperatures which cover the complete temperature range from $T
\ll J$ to $T \gg J$. Our results are for dimer like chains with
sufficiently small interdimer exchange $\lambda J$.  We find that the
characteristics of the noninteracting dimer limit (where the full
temperature dependent structure factor, Eq.~(\ref{eq:dynamicsfzero}),
is easily available analytically) describe much of the interacting
BAHC. This is demonstrated here for $\lambda = .3$ (but is essentially
true up to typically $\lambda \approx 0.5$) and applies in particular
to the existence of the central (zero frequency) peak and to the
temperature and wave vector dependence of the integrated intensities
('exclusive structure factors') for the central and the one magnon
peak. This large range of validity of the dimer picture is in
agreement with the observation \cite{Uhrig03} that the dimer picture
gives a good account of excitation strengths up to and including the
isotropic limit of the HAF at zero temperature.

Experimental and theoretical interest in the dynamics of the BAHC
results from the possibility that this theoretical model and its
realizations in a number of materials might serve as simple model
systems to discuss the interplay between temperature and quantum
fluctuations. Our results reveal that this interplay becomes apparent
in a number of points which are accessible to experimental, in
particular neutron scattering, observations:\\ (i) The shape of the
central peak which is a prominent feature of the BAHC, displays the
crossover with temperature from the noninteracting particle like
behavior at low temperatures to the diffusive behavior at high
temperatures.\\ (ii) The one magnon peak develops an asymmetric
linewidth (with appreciable strength on the high frequency side) with
temperature. This is particularly evident for wavevector $q=2 \pi$ at
temperatures $T \ge 1$ and appears to describe in more detail the
upward shift in gap energy with temperature noted in approximate
theoretical approaches.\\ (iii) The two magnon peak (around $q=\pi$)
is dominated by the bound triplet state on top of a small continuum which is
smooth at all temperatures and wavevectors.

The results as presented here are restricted to the basic situation of
equidistant spins and isotropy implying the absence of an external
magnetic field. In subsequent work we will extend the calculations to
cover the finite temperature dynamics in an external magnetic field
and also to describe real materials such as Cu(NO$_3$)$_2$ and
(VO)$_2$P$_2$O$_7$. We will also investigate in more detail the regime
$\lambda > 0.5$, approaching the isotropic (Heisenberg) limit in
order to study stronger deviations from the dimer like situation .\\

\section*{acknowledgments} 

We wish to thank M.~Enderle, A.~K.~Kolezhuk, S. Notbohm and D.~A.~Tennant for
stimulating discussions. We gratefully acknowledge that computational
facilities for the numerical calculations were generously
provided by the John von Neumann-Institut for Computing at J\"ulich
Research Center.


\begin{thebibliography}{50}
\bibitem{footnote01} We note that in our definition of reciprocal
  space the summation extends over the large Brillouin zone
  ($0<q<4\pi/b$), corresponding to the basic distance $d=b/2$ between
  spins.
\end{thebibliography}
\end{document}